\documentclass[12pt]{iopart}
\usepackage{graphicx} 
\newcommand{\ket}[1]{\mbox{$|\!#1\;\!\rangle$}}

\begin{document}
\title[Spin filling of a quantum dot derived from excited-state spectroscopy]{Spin filling of a quantum dot derived from excited-state spectroscopy}
\author{L.H. Willems van Beveren\footnote[3]{To whom correspondence should be addressed: laurens@qt.tn.tudelft.nl}, R. Hanson, I.T. Vink, \\ F.H.L. Koppens, L.P. Kouwenhoven and L.M.K. Vandersypen}
\address{Kavli Institute of Nanoscience, Delft University of Technology, \\PO Box 5046,
2600 GA Delft, The Netherlands}

\begin{abstract}
We study the spin filling of a semiconductor quantum dot using excited-state spectroscopy in a strong magnetic field. The field is oriented in the plane of the two-dimensional electron gas in which the dot is electrostatically defined. By combining the observation of Zeeman splitting with our knowledge of the absolute number of electrons, we are able to determine the ground state spin configuration for one to five electrons occupying the dot. For four electrons, we find a ground state spin configuration with total spin $S=1$, in agreement with Hund's first rule. The electron $g$-factor is observed to be independent of magnetic field and electron number.
\end{abstract}
\pacs{73.63.Kv, 03.67.Lx, 73.23.Hk, 71.70.Ej}
\submitto{New Journal of Physics, focus issue on ``Solid State Quantum Information''}

\section{Introduction}

A single electron spin confined in a semiconductor quantum dot (QD) is considered a promising candidate for the implementation of a qubit \cite{Los98,Vin00}.  
Also the joint spin state of $N$ electrons on a single dot may be used for quantum computation schemes. For instance, as long as the $N$-electron ground state has spin $S=1/2$, it can be used in a similar way as the spin of a single electron. Furthermore, proposals exist for encoding a qubit in two specific spin states of three electrons in a single dot, and for controlling this qubit fully electrically \cite{Kyr05}. Therefore, it is important to understand the interaction of multiple electron spins confined in a quantum dot, and specifically the spin configuration of the ground state. This can be done by studying spin filling, i.e. by determining the spin of successive electrons that are added to the dot, starting from zero electrons.

Quantum dots defined in pillars etched from a GaAs double-barrier heterostructure (``vertical'' QD) have  been studied extensively, showing spin filling obeying Hund's rule \cite{Tar96}, triplet-singlet ground state transitions \cite{Tar00a}, and a dependence of spin filling on the anisotropy of the  confinement potential \cite{Tar00b}. Furthermore, zero magnetic field addition spectra have revealed an atomic-like shell structure induced by a two-dimensional harmonic potential. Therefore these devices are commonly referred to as ``artificial atoms''. The addition spectrum of few-electron quantum dots, defined electrostatically within a two-dimensional electron gas (2DEG) by means of surface gates (so-called ``lateral" QDs), has also been studied \cite{Cio00}. Although a two-electron triplet-singlet ground state transition has been observed in these systems \cite{Kyr02}, evidence for a shell structure and spin filling obeying Hund's rule has not yet been found.

Here we study the spin filling of a few-electron lateral quantum dot by performing excited-state spectroscopy at a fixed magnetic field of 10 T, applied parallel ($B_{||}$) to the 2DEG. First we explain in detail our general method for determining spin filling. Then the device characteristics and settings are described. Finally we apply the method in order to determine spin filling for five successive transitions in the electron number, starting from an empty dot.

\section{Zeeman splitting and spin filling}

The method for determining spin filling is based on the facts that any single orbital can be occupied by at most two electrons, and that these electrons must have anti-parallel spins, due to the Pauli exclusion principle. Therefore, as we add one electron to a dot containing $N$ electrons, there are only two scenarios possible: (I) the electron moves into an empty orbital, or (II) it moves into an orbital that already holds one electron. We will now show that (in a high magnetic field) for the transition from the $N$-electron ground state, GS($N$), to the ($N$+1)-electron ground state, GS($N$+1), these two scenarios always correspond to the addition of a spin-up electron or a spin-down electron respectively. 

We first consider case I where an electron enters an empty orbital. In a strong magnetic field $B_{||}$, spin-up electrons have a lower energy than spin-down electrons \cite{spinfieldalignment} due to the Zeeman splitting $\Delta E_{Z} = g \mu_B B_{||}$, where $\mu_{B}$ = 58 $\mu$eV/T is the Bohr magneton. Therefore, if the orbital is empty, addition of a spin-up electron is energetically favored and thus takes the dot from GS($N$) to GS($N$+1). In contrast, addition of a spin-down electron takes the dot from GS($N$) to the ($N$+1)-electron excited state, ES($N$+1), which lies $\Delta E_{Z}$ higher in energy.

Next we look at case II, where an electron moves into an orbital with already one electron present. The electron that already occupies the orbital has spin-up if the dot is in GS($N$), as explained above. Therefore, the electron added in the transition from GS($N$) to GS($N$+1) must have spin-down in order to satisfy the Pauli exclusion principle. A spin-up electron can only be added to the same orbital if the first electron is spin-down, i.e. when the dot starts from ES($N$), $\Delta E_{Z}$ higher in energy than GS($N$). Thus, addition of a spin-up electron corresponds to a transition from ES($N$) to GS($N$+1). Comparing the two cases, we see that in case I, where a spin-up electron is added, there is an ($N$+1)-electron ES separated from GS($N$+1) by $\Delta E_{Z}$, while in case II, where a spin-down electron is added, there is a $N$-electron ES, separated from GS($N$) by $\Delta E_{Z}$. Thus, the spin filling has a one-to-one correspondence with the excited state spectrum.

\begin{figure}[t]
\begin{center}
\includegraphics[width=14cm, clip=true]{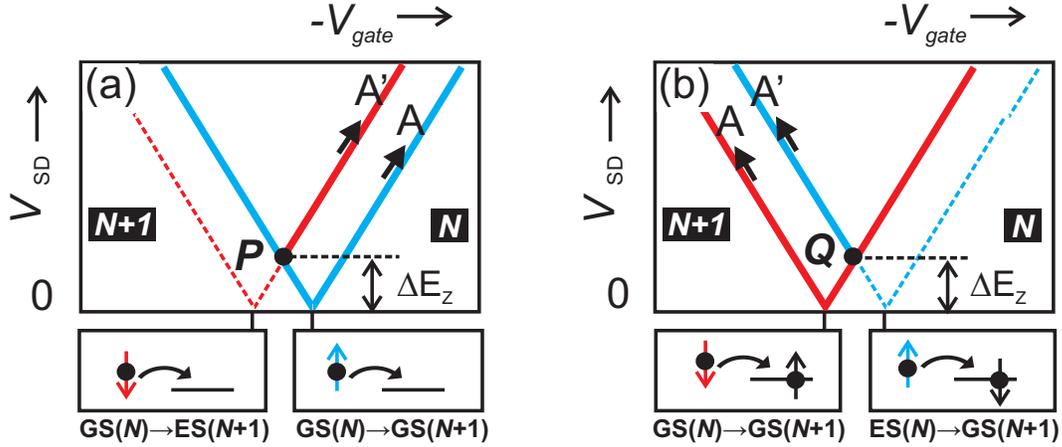}
\end{center}
\caption{\label{spinfilling} Schematic representation of excited-state spectra for case I (a) and case II (b). Shown is the differential conductance $dI/dV_{SD}$ as a function of $V_{SD}$ and gate voltage $V_{gate}$ in the presence of a strong in-plane magnetic field. The lines indicate where the transitions depicted in the corresponding diagrams become energetically accessible. Current can only flow in the V-shaped region defined by the GS($N$)-GS($N$+1) transition (line A); outside this region the dot is in Coulomb blockade and the number of electrons on the dot is fixed. When a transition involving an ES becomes accessible, the current changes, leading to an extra line in $dI/dV_{SD}$ (line A'). In case I, this line terminates at the ($N$+1)-electron CB region (point P in (a)), whereas in case II the line terminates at the $N$-electron CB region (point Q in (b)). Each of the excited-state spectra is symmetric with respect to $V_{SD}$. Therefore, the spectra for $V_{SD}<0$ can be obtained by rotating the shown spectra about the $V_{gate}$ axis.}
\end{figure}

We can discriminate between cases I and II by looking at electron transport through the dot as a function of the voltage bias ($V_{SD}$) applied between source and drain contacts, and gate voltage. \Fref{spinfilling} (a) and (b) schematically show the expected result of such a measurement for cases I and case II respectively. Lines in the differential conductance $dI/dV_{SD}$ indicate where electron transitions involving ground and excited states become energetically accessible. The transition from GS($N$) to GS($N$+1) is only allowed in the V-shaped region spanned by the two solid lines in $dI/dV_{SD}$ that intersect at $V_{SD} = 0$. These lines thus form the edges of the Coulomb blockaded (CB) region. The onset of the transition from GS($N$) to ES($N$+1), as in case I, appears as a line terminating at the edge of the ($N$+1)-electron CB region, at point $P$ in \fref{spinfilling}(a). In contrast, the onset of the transition from ES($N$) to GS($N$+1), as in case II, appears as a line terminating at the edge of the $N$-electron CB region, at point $Q$ in \fref{spinfilling} (b). 

Thus, if we see a line at a distance $\Delta E_{Z}$ from the edge of the CB region and terminating at the ($N$+1)-electron CB region, we have case I. Here a spin-up electron is added to the dot. In contrast, if there is a line at a distance $\Delta E_{Z}$ from the CB region that terminates at the $N$-electron CB region, we have case II, where a spin-down electron is added to the dot. 

The main requirement for this method is the ability to identify the Zeeman splitting in the excited-state spectrum. In GaAs lateral quantum dots Zeeman splitting has already been observed in several experiments \cite{Han03,Pot03,Kog04,Han04a}. We emphasize that the method is valid regardless of the spin $S$ of the ground states involved, as long as the addition of one electron changes the spin of the ground state by $\left|\Delta S\right|=1/2$ \cite{spin_blockade}. We now utilize this method to determine the change in spin at subsequent electron transitions in a lateral GaAs quantum dot containing zero to five electrons.

\section{Device characteristics}

\begin{figure}[t]
\begin{center}
\includegraphics[width=14cm, clip=true]{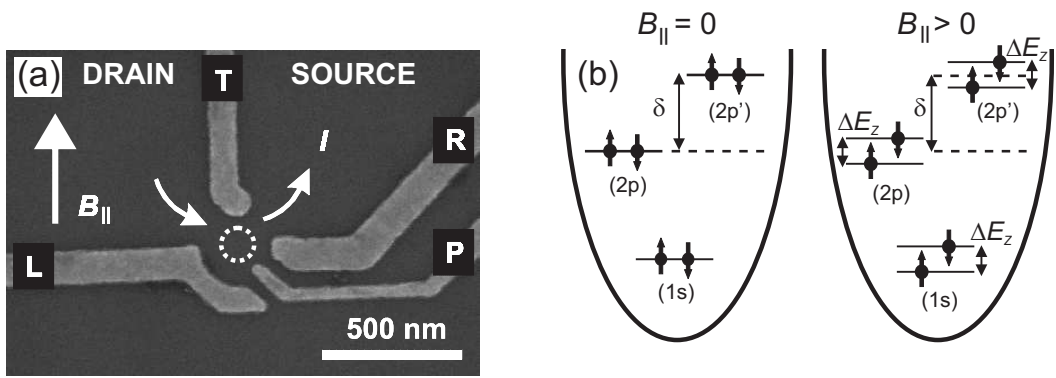}
\end{center}
\caption{(a) \label{figure_sample} Scanning electron micrograph of the device showing Ti/Au gate electrodes lying on top of a GaAs/AlGaAs heterostructure. The gate electrodes can locally deplete the 2DEG when negatively biased. Gate electrodes $L$, $R$ and $T$ are used to form the quantum dot, indicated by the white dashed circle. Gate electrode $P$ is left floating. (b) Level structure of a quantum dot with anisotropic parabolic confinement in zero and finite magnetic field. Two electrons with opposite spin can occupy the orbital states (1s), (2p) and (2p'). The latter two orbitals are separated in energy by $\delta$ due to  anisotropy of the confinement potential.}
\end{figure}

The lateral quantum dot is defined by Ti/Au gate electrodes patterned on top of a Si modulation doped GaAs/Al$_{x}$Ga$_{1-x}$As heterostructure ($x=0.265$), containing a high mobility 2DEG 60 nm below the surface, see \fref{figure_sample} (a). Ni/AuGe/Ni contacts are used to electrically connect to the source and drain reservoirs in the 2DEG. The 2DEG has an electron density ${n_{s}=4.0\times 10^{15}}$ m${^{-2}}$. This sample was cooled down with +266 mV on each of the surface gates in order to reduce background charge fluctuations \cite{Lad05}. The voltage on gate electrode $T$, $V_{T}$, is used to vary the electrochemical potential of the dot in each of the excited-state spectra. A magnetic field is applied parallel to the 2DEG to minimize additional lateral confinement and to exclude Landau level formation. All measurements are performed in a dilution refrigerator at base temperature $T=15$ mK. 

We tune the quantum dot to the few-electron regime at $B_{||}=0$ T. We identify the 0$\leftrightarrow$1 electron transition by the absence of further electron transitions in sweeping the gate voltages to more negative values under large applied source-drain voltage \cite{Kou01}. Then we track the 0$\leftrightarrow$1 electron transition as the magnetic field is swept to $B_{||}=10$ T. 

The charging energy of the dot, $E_{C}$, is 4.8 meV (for adding a second electron to a one-electron dot). For $N=1$, the level spacing from the orbital ground state to the first orbital excited state is 1.7 meV (data not shown). The level spacing between the first and second orbital excited state is considerably smaller, 0.8 meV. This implies that the confinement potential of the dot has no circular symmetry as in \cite{Foc30}. We nevertheless adopt the nomenclature from Refs. \cite{Kou01,Foc30} to denote the lowest orbital states in our quantum dot as (1s), (2p), etcetera. In fact, we shall see that the data is well explained by assuming an anisotropic confinement potential in the dot, where the two-fold orbital degeneracy of the first excited state (2p) is lifted. This gives rise to a level structure as shown in \fref{figure_sample} (b), in which the two (2p)-like orbitals, now denoted  (2p) and (2p'), are offset by an amount $\delta = 0.8$ meV. In lateral quantum dots, the spacing between successive orbitals is found to be dependent on gate voltage (and thus electron number) \cite{Kyr02,Kog03,Zum04}; generally, the level spacing decreases as the size of the dot is increased (which is needed to allow more electrons on the dot). Therefore, we also expect the value of $\delta$ to decrease as we increase the number of electrons in the dot.

We set the tunnel rate of the incoming barrier $\Gamma_{L}$ much smaller than the tunnel rate for the outgoing barrier $\Gamma_{R}$. As a result, in all of the excited-state spectra shown, the intensity of the lines involving transitions to or from excited states is enhanced when they run from bottom left to top right. In turn, the intensity of the lines involving excited states and running from top left to bottom right is suppressed \cite{Hay03}. Thus, lines corresponding to transitions from GS($N$) to ES($N$+1) (line A' in \fref{spinfilling} (a)) are most easily observed for $V_{SD}>0$, while lines corresponding to transitions from ES($N$) to GS($N$+1) (line A' in \fref{spinfilling} (b)) are most easily seen for $V_{SD}<0$.

\section{\textit{N}=0$\leftrightarrow$1 transition}

\begin{figure}[t]
\begin{center}
\includegraphics[width=16cm, clip=true]{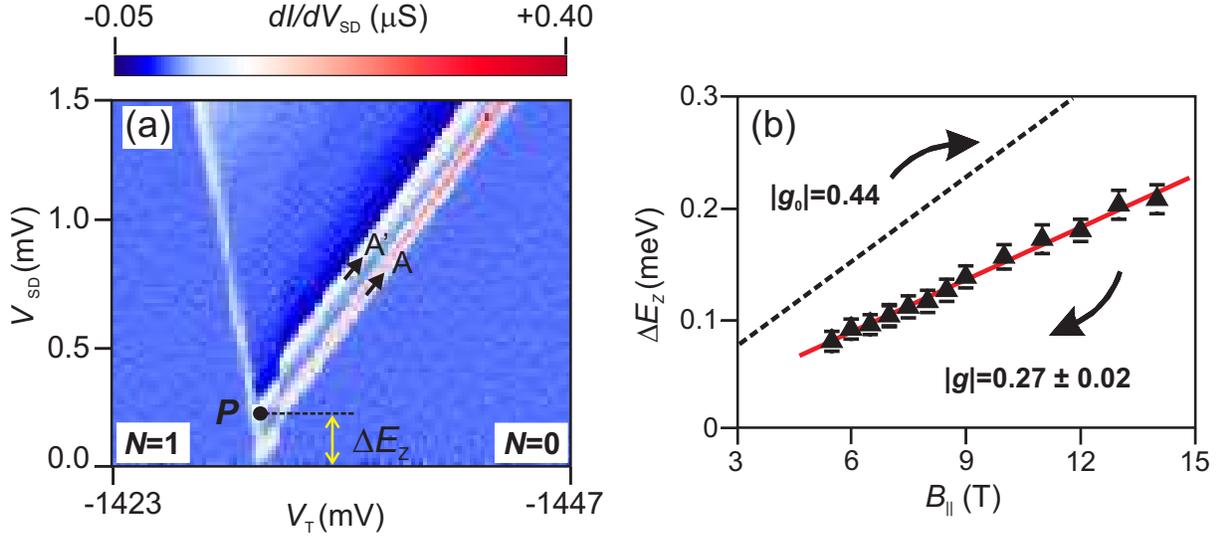}
\end{center}
\caption{(a) \label{figure_0_1} Color scale plot of the differential conductance $dI/dV_{SD}$ as a function of $V_{SD}$ and gate voltage $V_{T}$ near the 0$\leftrightarrow$1 electron transition, at $B_{||}=10$ T. Zeeman splitting of the orbital ground state is clearly observed (A-A'). The vertical shift in the data near $V_{T}$=-1430 mV is caused by a background charge rearrangement in the environment of the dot. (b) Extracted Zeeman splitting $\Delta E_{Z}$ at the 0$\leftrightarrow$1 electron transition as a function of $B_{||}$. A linear fit of $\Delta E_{Z}$ (red curve) results in $\left|g\right|= 0.27\pm0.02$. The dashed black line corresponds to the Zeeman splitting in bulk GaAs, where $g_{0}=-0.44$.}
\end{figure}

The excited-state spectrum obtained around the 0$\leftrightarrow$1 transition is shown in \fref{figure_0_1} (a). Clearly two parallel lines are observed, A and A'. The separation between these lines increases linearly with $B_{||}$, and  thus corresponds to the Zeeman splitting. From the spacing between lines A and A' (to be precise, from the value of $V_{SD}$ at point $P$), we extract $\Delta E_{Z}=0.16 \pm0.01$ meV at 10 T. Since A' terminates in the $N=1$ CB region, the electron added to the empty dot to form the $N=1$ GS has spin-up (see \fref{spinfilling} (a)), as expected.

In \fref{figure_0_1} (b) the Zeeman energy $\Delta E_{Z}$ at the 0$\leftrightarrow$1 transition is plotted versus applied magnetic field in the range from 5.5 to 14 T. For $B_{||}< 5.5$ T we cannot clearly resolve the Zeeman splitting from the spectroscopy data. As a reference, the Zeeman splitting $\Delta E_{Z}$ expected for bulk GaAs is plotted  with $\left|g_{0}\right|=0.44$ \cite{Wei77} (dashed black line). 
Clearly the $g$-factor we extract from \fref{figure_0_1} (b) is independent of field. Linear fitting of the data points results in a $g$-factor value of $0.27\pm0.02$. This value is lower than the bulk value of GaAs. Deviations of the $g$-factor from $g_{0}$ in quantum dots and possible explanations for this effect have been reported before \cite{Han03,Pot03,Kog04}.

\section{\textit{N}=1$\leftrightarrow$2 transition}

Next we tune the dot to the 1$\leftrightarrow$2 electron transition. At zero magnetic field the ground state for a two-electron quantum dot is always a spin singlet state \ket{\:S}, where two electrons with opposite spin occupy the lowest orbital and the total spin $S=0$ \cite{Ash74}. We expect that this is still true for an in-plane field of 10 T, as the Zeeman energy is much smaller than the zero-field singlet-triplet energy separation \cite{Han04a}. 

\begin{figure}[t]
\begin{center}
\includegraphics[width=16cm, clip=true]{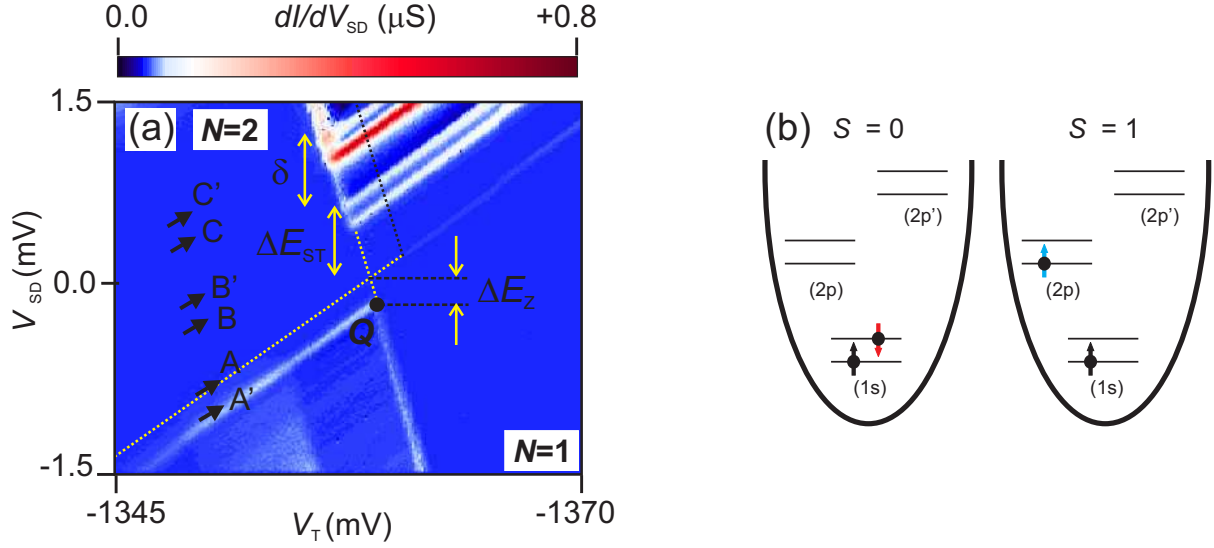}
\end{center}
\caption{(a) \label{figure_1_2} Color scale plot of the differential conductance $dI/dV_{SD}$ as a function of $V_{SD}$ and gate voltage $V_{T}$ near the 1$\leftrightarrow$2 electron transition, at $B_{||}=10$ T. (b) \textit{Left}: Ground state spin configuration for a two-electron dot. A spin-down electron (red) is added and pairs with the spin-up electron already present, to form a singlet state \ket{\:S} with total spin $S=0$. \textit{Right}: Spin configuration of an excited state. The second spin-up electron occupies the (2p) orbital to form a two-electron triplet state \ket{\:T} with total spin $S=1$. When the second spin-up electron occupies the (2p') orbital an excited triplet state is formed.}
\end{figure}

The $dI/dV_{SD}$ data obtained for the 1$\leftrightarrow$2 electron transition is shown in \fref{figure_1_2} (a). Two lines A and A' separated by $0.16 \pm0.01$ meV are visible, for $V_{SD}<0$. This is exactly the energy scale of the Zeeman splitting found for the 0$\leftrightarrow$1 transition. Because A' terminates in $Q$ at the edge of the $N=1$ CB region (and not at the edge of the $N=2$ CB region), we conclude that the transition from GS(1) to the GS(2) involves adding a spin-down electron. The spin-down electron pairs with the spin-up electron already present in the (1s) orbital to form a two-electron singlet state \ket{\:S}, as illustrated in the left diagram in \fref{figure_1_2} (b).

For the color scale chosen, parts of line A are difficult to observe. Therefore dashed yellow lines are added  as a guide to the eye. For $V_{SD}>0$, line A' is also hardly visible. Its position is indicated by a black dashed line. The absence of this latter line is caused by the asymmetry of the tunnel barriers, as explained earlier. For the same reason, the lines B-B' and C-C' are hardly visible for $V_{SD}<0$. Line A is weaker than line A' because in a strong magnetic field spin-up electrons generally couple better to the source and drain reservoirs than spin-down electrons, even if the magnetic field is applied in the plane of the 2DEG \cite{Han04b}.

The lines B-B' correspond to the onset of transitions involving the three triplet states. These lines too are separated by the Zeeman splitting; B and B' involve transport of spin-up and spin-down electrons respectively \cite{Han04a}. The distance between A and B' gives the singlet-triplet energy splitting $E_{ST}$, $0.56 \pm0.02$ meV. The data also shows a third set of parallel lines with the same spacing, C-C'. The fact that the pair of lines C-C' have a different intensity than the pair of lines B-B' suggests that different orbitals are involved. We believe that the lines C-C' correspond to transitions to and from triplet states with one electron in the (2p') orbital instead of in the (2p) orbital. Here, the offset between the (2p) and (2p') orbitals, $\delta$, is 0.52 meV (B'-C'), somewhat smaller than the value at the 0$\leftrightarrow$1 electron transition. 

\section{\textit{N}=2$\leftrightarrow$3 transition}

Next we move on to the 2$\leftrightarrow$3 electron transition. We have seen that the two-electron ground state is a singlet state \ket{\:S}, where two electrons with opposite spin occupy the (1s) orbital. When a third electron is added, it has to occupy a next orbital in order to satisfy the Pauli exclusion principle. 

\begin{figure}[t]
\begin{center}
\includegraphics[width=16cm, clip=true]{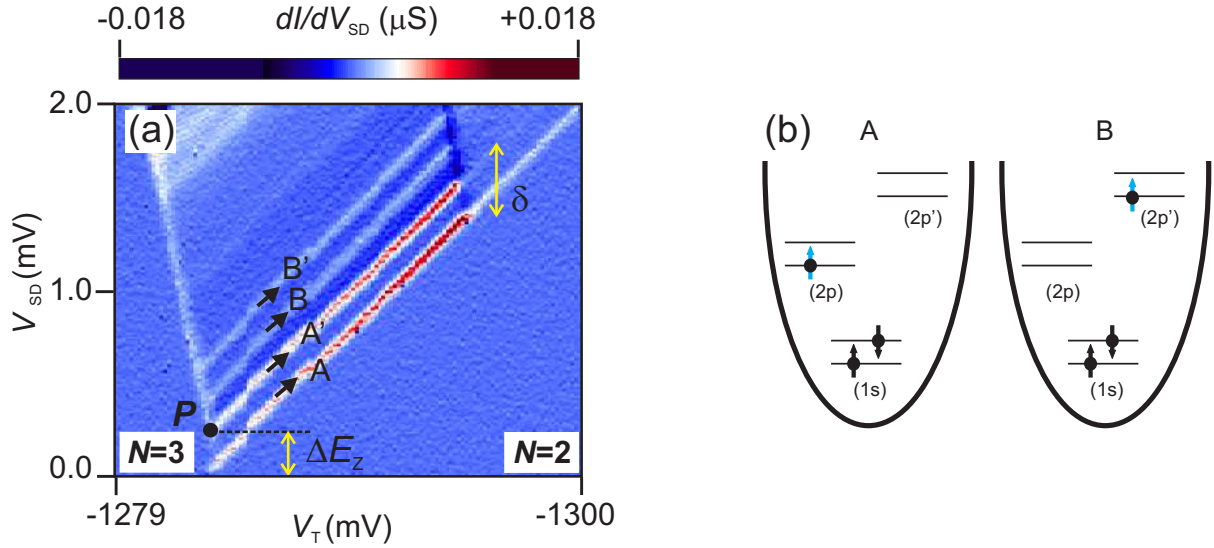}
\end{center}
\caption{(a) \label{figure_2_3} Color scale plot of the differential conductance $dI/dV_{SD}$ as a function of $V_{SD}$ and gate voltage $V_{T}$ near the 2$\leftrightarrow$3 electron transition, at $B_{||}=10$ T. (b) \textit{Left}: Ground state spin configuration for a three-electron dot. \textit{Right}: Spin configuration of the excited state that can be reached starting at line B in (a).}
\end{figure}

The $dI/dV_{SD}$ data we find is shown in \fref{figure_2_3} (a). We notice several important features. A first pair of lines, A-A', is split by $0.17\pm0.01$ meV, the Zeeman splitting. Since these lines terminate at the edge of the $N=3$ CB region, the transition from GS(2) to GS(3) involves adding a spin-up electron. The three-electron ground state then corresponds to the situation where the third electron (with spin-up) occupies the orbital (2p), as shown in the left diagram of \fref{figure_0_1} (b). 

A second set of Zeeman split lines, B-B', runs parallel to the lines A and A', with smaller amplitude. The separation between both sets of transitions, $\delta$, is 0.34 meV. In our spin filling picture, the lines B and B' correspond to electron transitions where the third electron occupies the (2p') orbital, as in the right diagram of  \fref{figure_0_1} (b). The value of $\delta$ we find here is smaller than at the 1$\leftrightarrow$2 electron transition, as expected from our earlier considerations. 

\section{\textit{N}=3$\leftrightarrow$4 transition}

\begin{figure}[t]
\begin{center}
\includegraphics[width=16cm, clip=true]{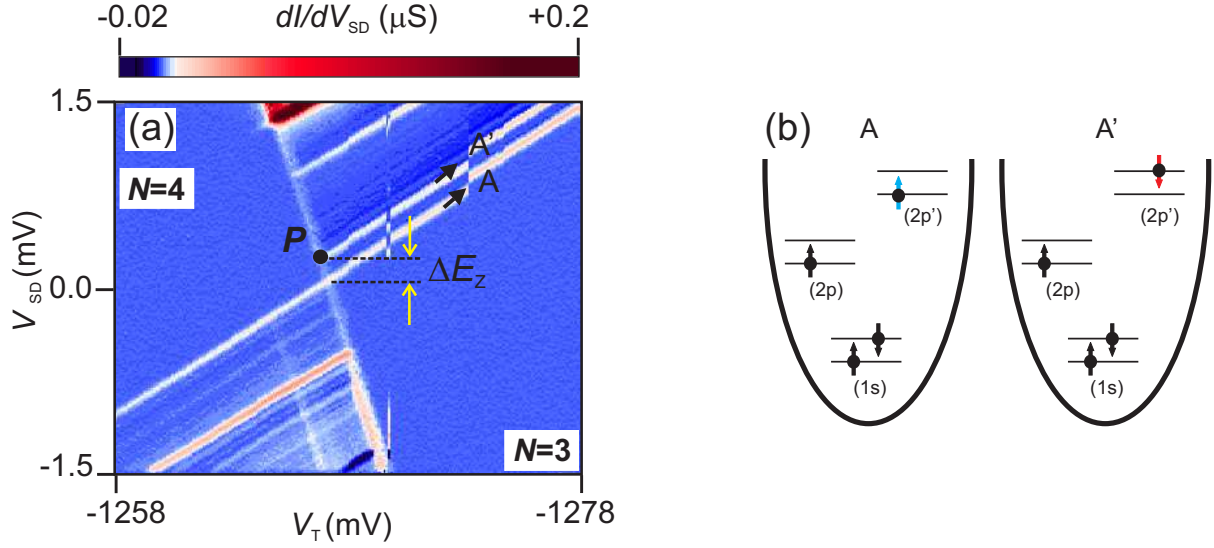}
\end{center}
\caption{(a) \label{figure_3_4} Color scale plot of the differential conductance $dI/dV_{SD}$ as a function of $V_{SD}$ and gate voltage $V_{T}$ near the 3$\leftrightarrow$4 electron transition, at $B_{||}=10$ T. (b) \textit{Left}: Ground state spin configuration of the four-electron dot ($S=1$). \textit{Right}: Spin configuration of one possible spin-excited state, where the (2p') orbital is occupied by a spin-down electron.}
\end{figure}

In \fref{figure_3_4} (a) we show the excited-state spectrum for the 3$\leftrightarrow$4 electron transition. Ignoring spin-exchange interactions, one expects a four-electron dot with total spin $S=0$, with two electrons in the (1s) orbital and two electrons in the (2p) orbital. This implies that a spin-down electron must be added to a three-electron dot in the GS in order to reach the $N=4$, $S=0$ state.

However, the data in \fref{figure_3_4} (a) indicates that the fourth electron added to the dot has spin-up, because the line A', separated from line A by the Zeeman splitting ($\Delta E_{Z}=0.17\pm0.01$ meV), terminates at the edge of the $N=4$ CB region. Since both the third and fourth electron have spin-up, they occupy different orbitals, (2p) and (2p'), as shown in the left diagram of \fref{figure_3_4} (b). The $N=4$ ground state thus has $S=1$. This $N=4$ ground state can be understood when we take into account the exchange interaction $K_{ab}$ between the spins in the (2p) and (2p') orbital and the terms $C_{aa}$ and $C_{ab}$, representing the direct Coulomb energy when the two spins are in the same or in a different orbital state respectively \cite{Kou01}. $S=1$ spin filling is favored when $K_{ab}+\left|C_{aa}-C_{ab}\right|>\delta$.

This particular spin configuration of the $N=4$ ground state is related to Hund's first rule, which states that a shell of degenerate orbitals will, as much as possible, be filled by electrons with parallel spins, up to the point where the shell is half filled. Exchange energy $K_{ab}$ (causing a lowering of the Coulomb energy when spins align parallel in different orbitals) reduces the total energy and favors the $S=1$ state. This state can only exist when the non-degeneracy is small compared to the exchange energy and the difference in direct Coulomb energy terms \cite{Tar00b}.

\section{\textit{N}=4$\leftrightarrow$5 transition}

As we have seen, the four-electron ground state has a total spin $S=1$, in agreement with Hund's first rule. When the fifth electron is added the total spin of the system is expected to change back from $S=1$ to $S=1/2$ because of spin pairing in the (2p) orbital. A less likely option for the transition from GS(4) to GS(5) is that the fifth electron tunnels into the next empty orbital.

\begin{figure}[t]
\begin{center}
\includegraphics[width=16cm, clip=true]{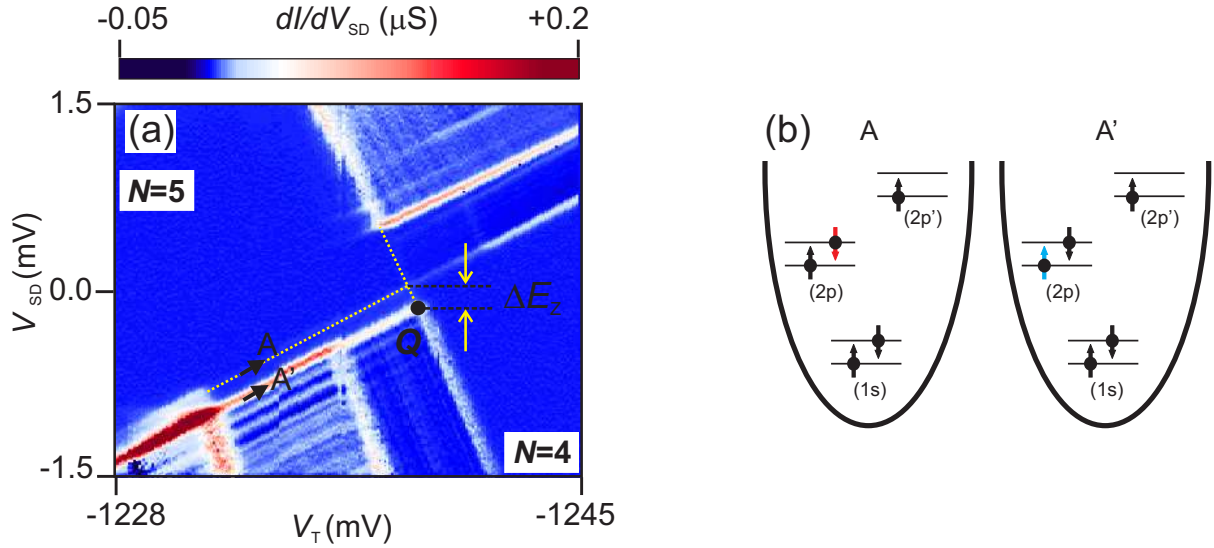}
\end{center}
\caption{(a) \label{figure_4_5} Color scale plot of the differential conductance $dI/dV_{SD}$ as a function of $V_{SD}$ and gate voltage $V_{T}$ near the 4$\leftrightarrow$5 electron transition, at $B_{||}=10$ T. (b) \textit{Left}: Ground state spin configuration of a five-electron dot.  \textit{Right}: Spin configuration of the spin-excited state. In both cases, the total spin is $S=1/2$.}
\end{figure}

In \fref{figure_4_5} (a) we show the excited-state spectrum for the 4$\leftrightarrow$5 electron transition. Indeed, the transition from GS(4) to GS(5) corresponds to adding an electron with spin-down, as the line A' terminates at the edge of the $N=4$ CB region (as before, some of the lines are hardly visible; their position is indicated by yellow dashed lines). The resulting spin configuration for five electrons is indicated in the left diagram of \fref{figure_4_5} (b). The added spin-down electron occupies the (2p) orbital. The Zeeman splitting extracted from the data ($V_{SD}<0$) is $\Delta E_{Z}=0.17\pm0.01$ meV.

\section{Conclusion}

We have determined the spin filling of a few-electron lateral quantum dot containing one up to five electrons in a parallel magnetic field of 10 T. The spin filling was extracted without magnetic field sweeps, by looking at the position of the excited spin state in the spectroscopy data. The Zeeman splitting is equal for all orbitals and independent of the number of electrons on the dot. The ground state of a four-electron dot is a $S=1$ Hund state. Field dependence of the Zeeman splitting for the 0$\leftrightarrow$1 electron transition yields a field independent $g$-factor value of $0.27\pm 0.02$. 

\ack
This work was supported by the Dutch Science Foundation NWO/FOM, the DARPA-QUIST program, the ONR, the International Cooperative Research Project (ICORP) and the EU-RTN network on spintronics.


\section*{References}

\begin{thebibliography}{99}
\bibitem{Los98} D. Loss and D. P. DiVincenzo, Phys.\ Rev.\ A {\bf 57}, 120 (1998).
\bibitem{Vin00} D. P. DiVincenzo, Nature {\bf 408}, 339 (2000).  
\bibitem{Kyr05} J. Kyriakidis \etal, Phys.\ Rev.\ B {\bf 71}, 125332 (2005).
\bibitem{Tar96} S. Tarucha \etal, Phys.\ Rev.\ Lett.\ {\bf 77}, 3613 (1996).
\bibitem{Tar00a} S. Tarucha \etal, Phys.\ Rev.\ Lett.\ {\bf 84}, 2485 (2000).
\bibitem{Tar00b} S. Tarucha \etal, Appl.\ Phys.\ A {\bf 71}, 367 (2000).
\bibitem{Cio00} M. Ciorga \etal, Phys.\ Rev.\ B {\bf 61}, R16315 (2000).
\bibitem{Kyr02} J. Kyriakidis \etal, Phys.\ Rev.\ B {\bf 66}, 035320 (2002).
\bibitem{spinfieldalignment} 
Since the $g$-factor in GaAs is negative, the electron spin is aligned with the magnetic moment of the electron, and spin-up (aligned with the magnetic field) has a lower energy than spin-down. 
\bibitem{Han03} R. Hanson \etal, Phys.\ Rev.\ Lett.\ {\bf 91}, 196802 (2003).
\bibitem{Pot03} R. M. Potok \etal, Phys.\ Rev.\ Lett.\ {\bf 91}, 016802 (2003).
\bibitem{Kog04} A. Kogan \etal, Phys.\ Rev.\ Lett.\ {\bf 93}, 166602 (2004).
\bibitem{Han04a} R. Hanson \etal, Phys.\ Rev.\ B {\bf 70}, 241304 (2004).
\bibitem{spin_blockade}
If $|\Delta S| > 1/2$, the ($N$+1)-electron GS cannot be reached from the $N$-electron GS by adding a single electron, leading to a spin blockade of electron transport through the dot, see D. Weinmann \etal, Phys.\ Rev.\ Lett.\ {\bf 74}, 984 (1995).
\bibitem{Lad05} M. Pioro-Ladri\`{e}re \etal, cond-mat/0503602.
\bibitem{Kou01} L. P. Kouwenhoven \etal, Rep.\ Prog.\ Phys.\ {\bf 64} (6), 701 (2001).
\bibitem{Foc30} C. G. Darwin, Proc.\ Cambridge.\ Philos.\ Soc.\ {\bf 27}, 86 (1930); V. Fock, Z.\ Phys.\ {\bf 47}, 446 (1928).
\bibitem{Kog03} A. Kogan \etal, Phys.\ Rev.\ B {\bf 67}, 113309 (2003).
\bibitem{Zum04} D. M. Zumb\"{u}hl \etal, Phys.\ Rev.\ Lett.\ {\bf 93}, 256801 (2004).
\bibitem{Hay03} T. Hayashi \etal, Phys.\ Stat.\ Sol.\ B {\bf 238}, 262 (2003).
\bibitem{Wei77} C. Weisbuch and C. Hermann, Phys.\ Rev.\ B {\bf 15}, 816 (1977).
\bibitem{Ash74} N. W. Ashcroft and N. D. Mermin, {\it Solid State Physics, New York: Saunders} (1974).
\bibitem{Han04b} R. Hanson \etal, {\it Proceedings of the 39th Rencontres de Moriond}, see also cond-mat/0407793.

\end{thebibliography}
\end{document}